\documentclass[dvips,11pt,a4paper,twoside]{article}
\usepackage[spanish]{babel} 
\usepackage{amsmath}
\usepackage[dvips]{graphics}
\usepackage{graphicx}
\numberwithin{equation}{section}
\numberwithin{figure}{section}
\numberwithin{table}{section}

\usepackage{color}

\definecolor{lightgray}{gray}{0.5}
\title{Numerical response of the magnetic permeability as a
funcion of the frecuency of NiZn ferrites using Genetic Algorithm}

\author{Silvina Boggi, Adrian C. Razzitte,Gustavo Fano} 

\begin{document}
\maketitle

\begin{abstract}
The magnetic permeability of a ferrite is an important factor in designing devices such as inductors, transformers, and microwave absorbing materials among others. Due to this, it is advisable to study the magnetic permeability of a ferrite as a function of frequency.

When an excitation that corresponds to a harmonic magnetic field \textbf{H} is applied to the system, this system responds with a magnetic flux density \textbf{B}; the relation between these two vectors can be expressed as \textbf{ B }=$\mu(\omega)$ \textbf{H}  . Where $\mu$ is the magnetic permeability.

In this paper, ferrites were considered linear, homogeneous, and isotropic materials. A magnetic permeability model was  applied to NiZn ferrites doped with Yttrium. 

The parameters of the model were adjusted using the Genetic Algorithm. In the computer science field of artificial intelligence,  Genetic Algorithms and Machine Learning does rely upon nature's bounty for both inspiration nature's and mechanisms. Genetic Algorithms are probabilistic search procedures which generate solutions to optimization problems using techniques inspired by natural evolution, such as inheritance, mutation, selection, and crossover.

For the numerical fitting  usually is used a nonlinear least square method, this  algorithm  is  based on calculus  by starting from an initial set of variable values. This approach is mathematically elegant compared to the exhaustive or random searches but  tends  easily to get stuck in local minima. On the other hand, random methods use some probabilistic calculations to find variable sets. They tend to be slower but have greater success at finding the global minimum regardless of the initial values of the variables

\end{abstract}



\section{Magnetic permeability model}

The ferrites materials have been widely used as various electronic devices such as inductors, transformers, and electromagnetic wave absorbers in the relatively high-frequency region up to a few hundreds of MHz.

The electromagnetic theory can be used to describe the macroscopic properties of matter. The electromagnetic fields may be characterized by four vectors: electric field \textbf{E}, magnetic flux density \textbf{B }, electric flux density \textbf{D}, and magnetic field \textbf{H}, which at ordinary points satisfy Maxwell's equations.

The ferrite media under study  can be considerer as linear, homogeneous, and isotropic. The relation between the vectors \textbf{B} and \textbf{H } can be expressed as : \textbf{B }=$\mu(\omega)$\textbf{H }. Where $\mu$ is the magnetic permeability of the material.
 
Another important parameter for magnetic materials is magnetic susceptibility $\chi$ which relates the magnetization vector M to the magnetic field vector \textbf{H} by the relationship: \textbf{M }=$\chi(\omega)$ \textbf{H }.

Magnetic permeability $\mu$ and magnetic susceptibility $\chi$  are related by the formula: $\mu=1 + \chi $.

Magnetic materials in sinusoidal fields have, in fact, magnetic losses and this can be expressed taking $\mu$  as a complex parameter:$ \mu=\mu' +  j  \mu" $ \cite{VonHippel}

In the frequency range from RF to microwaves, the complex permeability spectra of the ferrites can be characterized by two different magnetization mechanisms: domain wall motion and gyromagnetic spin rotation.

Domain wall motion contribution to susceptibility can be studied through an equation of motion in which pressure is proportional to the magnetic field \cite{greiner}.

Assuming that the magnetic field has harmonic excitation $H= H_{0} e^{j\omega t}$ ,  the contribution of domain wall to the susceptibility $\chi _{d}$ is:

\begin{equation}\label{eq:chid}
\chi_{d}=\frac{\omega^{2}\;\chi_{d0}}{\omega^{2}{_{d}}-\omega^{2}-j\omega\beta}
\end{equation}\ 

Here, $\chi_{d}$ is the magnetic susceptibility for domain wall, $\omega_{d}$ is the resonance frequency of domain wall contribution,  $\chi_{d0}$ is the static magnetic susceceptibility, $\beta$ is the damping factor and $\omega$ is the frequency of the external magnetic field.

\vspace{0.2cm}

 Gyromagnetic spin contribution to magnetic susceptibility can be studied through a magnetodynamic equation \cite{sohoo}\cite{wohlfarth}.

The magnetic susceptibility $\chi_{s}$ can be expressed as:

\begin{equation}\
\chi_{s}=\frac{\left(\omega_{s}-j\omega\alpha\right)\omega_{s}\chi_{s0}}{\left(\omega_{s}-j\omega\alpha\right)^{2}-\omega^{2}},
\end{equation}\

Here,  $\chi_{s}$ is the magnetic susceptibility for  gyromagnetic spin, $\omega_{s}$ is the resonance frequency of spin contribution,   $ \chi_{s0}$ is the static magnetic susceptibility,  and $\alpha$  is the damping factor and $\omega$ is the frequency of the external magnetic field.

The total magnetic permeability results \cite{PhysicaB}:

\begin{equation}\label{eq:modelo}
\mu=1+ \chi_{d}+\chi_{s}=1+\frac{\omega^{2}\;\chi_{d0}}{\omega^{2}{_{d}}-\omega^{2}-j\omega\beta}+\frac{\left(\omega_{s}+j\omega\alpha\right)\omega_{s}\chi_{s0}}{\left(\omega_{s}+j\omega\alpha\right)^{2}-\omega^{2}}
\end{equation}\

Separating the real and the imaginary parts of equation (\ref{eq:modelo}) we get:

\begin{equation}\label{mureal}
\mu'\left(\omega\right)=1+\frac{\omega_{d}^{2}\;\chi_{d0}\left(\omega_{d}^{2}-\omega^{2}\right)}{\left(\omega_{d}^{2}-\omega^{2}\right)^{2}+\omega^{2}\beta^{2}}+\frac{\omega_{s}^{2}\;\chi_{s0}\left(\omega_{s}^{2}-\omega^{2}\right)+\omega^{2}\alpha^{2}}{\left(\omega_{s}^{2}-\omega^{2}\left(1+\alpha^{2}\right)\right)^{2}+4\omega^{2}\omega_{s}^{2}\alpha^{2}}
\end{equation}

\begin{equation}\label{muimag}
\mu"\left(\omega\right)=1+\frac{\omega_{d}^{2}\;\chi_{d0}\;\omega\;\beta}{\left(\omega_{d}^{2}-\omega^{2}\right)^{2}+\omega^{2}\beta^{2}}+\frac{\omega_{s}\;\chi_{s0}\;\omega\;\alpha\left(\omega_{s}^{2}+\omega^{2}\left(1+\alpha^{2}\right)\right)}{\left(\omega_{s}^{2}-\omega^{2}\left(1+\alpha^{2}\right)\right)^{2}+4\omega^{2}\omega_{s}^{2}\alpha^{2}},
\end{equation}

Magnetic losses, represented by the imaginary part of the magnetic permeability, can be extremely small; however, they are always present unless we consider vacuum \cite{landau}.
From a physics point of view, the existing relationship between $\mu'$ and $\mu"$ reflects that the mechanisms of energy storage and dissipation are two aspects of the same phenomenon \cite{boggi}.

\section{Genetic Algorithms}

Genetic Algorithms (GA) are probabilistic search procedures which generate solutions to optimization problems using techniques inspired by natural evolution, such as inheritance, mutation, selection, and crossover.
 
A GA allows a population composed of many individuals evolve according to selection rules designed to maximize "`fitness"' or minimize a "`cost function"'.

\begin{figure}
	\centering
		\includegraphics[width=0.7\textwidth]{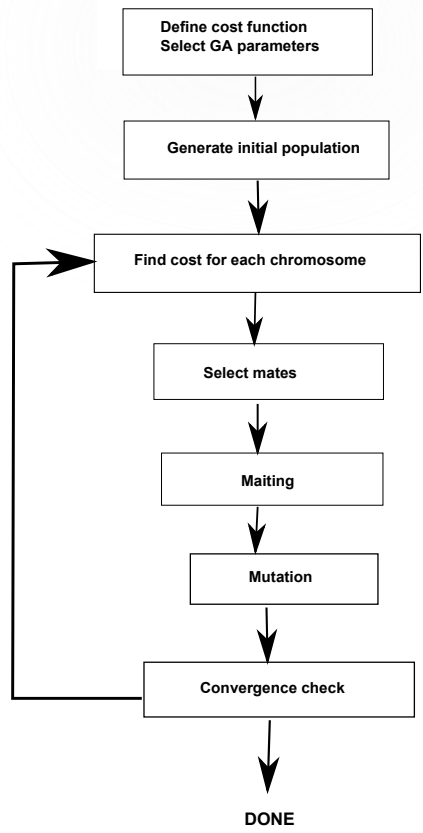}
	\caption{Flowchart of a  Genetic Algorithm.}
	\label{fig:diagramaflujo}
\end{figure}

A path through the components of AG is shown as a flowchart in Figure (\ref{fig:diagramaflujo})

\subsection{Selecting the Variables and the Cost Function}

A cost function generates an output from a set of input variables (a chromosome). The  Cost function's object is to modify  the output in some desirable fashion by finding the appropriate values for the input variables.  The Cost function in this work is the difference between the experimental  value of the permeability and calculated using the parameters obtained by the genetic algorithm.

To begin the AG is randomly generated an initial population of chromosomes. This population is represented by a matrix in which each row is a chromosome that contains the variables to optimize, in this work, the  parameters of permeability model. \cite {1}

\subsection{Natural Selection }

Survival of the fittest translates into discarding the chromosomes with the highest cost . First, the  costs and associated chromosomes are ranked from lowest cost to highest cost. Then, only the best are selected to continue, while the rest are deleted. The selection rate, is the fraction of chromosomes that survives for the next step of mating.

\subsection{ Select mates}

Now two chromosomes are selected from the set surviving to produce two new offspring which  contain traits from each parent. 
Chromosomes with lower cost are more likely to be selected from the chromosomes that survive natural selection. Offsprings are born to replace the discarded chromosomes

\subsection{Mating}

The simplest methods choose one or more points in the chromosome to
mark as the crossover points. Then the variables between these points are
merely swapped between the two parents. 
Crossover points are randomly selected.

\subsection{Mutaci\'on}

If care is not taken, the GA can converge too quickly into one  region of a local minimum of the cost function rather than a global minimum. To avoid this problem of overly fast convergence, we force the routine to explore other areas of the cost surface by randomly introducing changes, or mutations, in some of the
variables.

\subsection{ The Next Generation}

The process described is iterated until an acceptable solution is found.
The individuals of the new generation (selected, crossed and mutated)  repeat the whole process until it reaches a termination criterion. In this case, we consider a maximum number of iterations or a predefined acceptable solution (whichever comes first)

\section{Results and discussion }

$Ni_{0.5}Zn_{0.5}Fe_{2-x}Y_{x}O_{4}$ samples were prepared via sol-gel method with x=0.01,
0.02, and 0.05.  The complex permeability of the samples was measured in a material analyzer
HP4251 in the range of 1MHz to 1 GHz.\cite{silvia}.

The experimental data of magnetic permeability have been used for fitting the parameters of the model \cite{PhysicaB}. Firstly, we fitted the magnetic losses based in  equation (\ref{muimag}) by the Genetic Algoritm method  and we obtained the six fitting parameters. 
We substituted, then,  these six parameters  into equation (\ref{mureal}) to calculate the real part of permeability.

The variables of the problem to adjust were the six parameters of the model: ($\chi_{d}$,  $\chi_{s}$, $\omega_{d} $, $\omega_{s}$, $\beta$ y $\alpha$) 

Magnetic losses being a functional relationship: 
\begin{equation}
\mu''_{ajustado}= f(\omega,\chi_{d},\chi_{s},\omega_{d},\beta,\alpha)	
\end{equation}
	
where $\omega$ is the frequency of the external magnetic field, y $\chi_{d}$,  $\chi_{s}$, $\omega_{d} $, $\omega_{s}$, $\beta$   $\alpha$ are unknown parameters, the problem is to estimate these from a set of pairs of experimental:$\:\:(\omega_{i},\mu''_{i}); (i=1,2,...,n)$ .

The cost function was the mistake made in calculating $\mu''_{ajustado}$  with the expression (\ref {muimag}) using the parameters obtained from the genetic algorithm and the experimental value of $\mu''$ for each frequency.

	\[Cost\;function=\sum^{n}_{i=1}(f(\omega_{i},\chi_{d},\chi_{s},\omega_{d}, \omega_{s}, \beta,\alpha)-\mu''_{i})^{2}
\]

2000 iterations were performed, with a population of 300 chromosomes (each with 6 variables). The fraction of the population that was replaced by children in each iteration was 0.5 and the fraction of mutations was 0.25.

\begin{figure}
		\includegraphics[width=0.8\textwidth]{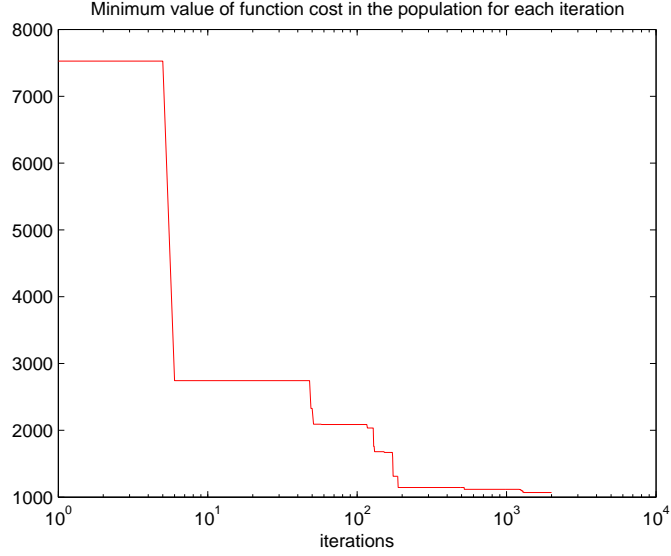}
	\label{fig:mincost}
	\caption{ Evolution of the error}
\end{figure}

The figure \ref{fig:mincost} shows the evolution of the error in successive iterations, we graphed the minimum value of function cost in the population for each iteration. It show that  the error converges at minimum value quickly,  and then this  value is stable.

Although in the equations (\ref{muimag}) and (\ref{mureal}),  $\omega_{d}$ and $\omega_{s}$ must be in Hz, we calculate in MHz and then  multiplied in the equations by $1.10^{6}$, the same treatment we performed with the parameter beta, we calculate its value in the range between 1 and 2000, although in equations we multiplied  by $1.10^{7}$. This treatment was necessary for that  the AG values to variables within a limited range.

\begin{figure}h
	\includegraphics[width=1\textwidth]{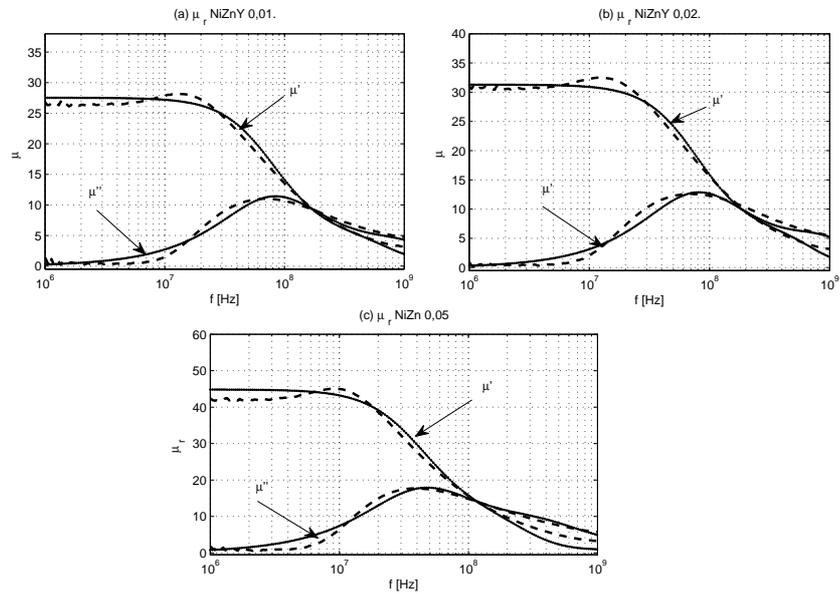}
	\caption{(a)(b)y(c) Complex permeability spectra in NiZn ferrites}
	\label{fig:permeajuste}
\end{figure}

Table \ref{tab:tabla1} shows the parameters of the model calculated for the three NiZn ferrite samples doped with different amounts of Yttrium.
Figure \ref{fig:permeajuste} graphs (a), (b) and (c) shows the permeability spectra for the three ferrites. Solid lines represent the curves constructed from the adjusted parameters while dotted and dashed lines represent the curves from experimental data.

The frequency of the  $\mu''$ maximum for the spin component is calculated \cite{tsutaoka}:

\begin{equation}
\omega_{\mu''\!max}^{s}=\frac{\omega_{s}}{\sqrt{1+\alpha^{2}}}
\end{equation}
And for the domain wall component:\cite{tsutaoka}:

\begin{equation}
\omega_{\mu''\!max}^{d}= \frac{1}  {6}\sqrt{12\omega_{d}^{2}-6\beta^{2}+6\sqrt{16\omega_{d}^{4}-4\omega_{d}^{2}\beta^{2}+\beta^{4}}}
\end{equation}

In these ferrites maximums are located in: $\omega_{\mu''\!max}^{s}\cong\!80\: MHz$ y  $\omega_{\mu''\!max}^{d}\cong\!1000\: MHz$. 

\begin{table}[htb] 
\centering
\begin{tabular}{|c|c|c|c|c|c|c|}
\hline
& $\chi_{d0}$ & $\omega_{d}\;\left(Hz\right)$ & $\beta$ & $\chi_{s0}$ & $\omega_{s}\;\left(Hz\right)$ & $\alpha$\\
\hline
\hline
 NiZnY 0.01 & 22.05 &1262 $\cdot10^{6}$ & 1966$\cdot10^{7}$ & 4.48 &1989$\cdot10^{6}$ & 1.8967 \\
\hline
 NiZnY 0.02 & 24.79 & 1115$\cdot10^{6}$ & 1581$\cdot10^{7}$ & 5.50 & 1480$\cdot10^{6}$ & 1.40 \\
\hline
 NiZnY 0.05 & 33.75 & 671$\cdot10^{6}$ & 1021$\cdot10^{7}$ & 10.75 & 1334$\cdot10^{6}$ & 3.477 \\
\hline
\end{tabular}
\caption{Adjusted parameters in the permeability model for NiZn ferrites doped with Yttrium.}
\label{tab:tabla1}
\end{table} 

\end{document}